\documentclass[runningheads]{svmult}

\usepackage{makeidx}   
\usepackage{graphicx}  
\usepackage{subeqnar}  
\usepackage{multicol}  
\usepackage{physprbb}  
\makeindex             

\begin{document}
\title*{Multiplicity of young brown dwarfs in Cha\,I}
\toctitle{Focusing of a Parallel Beam to Form a Point
\protect\newline in the Particle Deflection Plane}
%
%
\titlerunning{Multiplicity of young brown dwarfs in Cha\,I}
%
\author{Viki Joergens\inst{1}
\and Eike Guenther\inst{2}
\and Ralph Neuh\"auser\inst{1}
\and Fernando Comer\'on\inst{3}
\and Nuria Hu\'elamo\inst{1}
\and Jo\~ao Alves\inst{3}
\and Wolfgang Brandner\inst{3}
}
\authorrunning{Viki Joergens et al.}
%
%
\institute{Max-Planck-Institut f\"ur Extraterrestrische Physik, 
	Giessenbachstr. 1,
	D-85748 Garching, 
	Germany
\and Th\"uringer Landessternwarte Tautenburg,
	Karl-Schwarzschild-Observatorium,
	Sternwarte 5
	D-07778 Tautenburg,
	Germany   
\and European Southern Observatory
	Karl-Schwarzschild-Str. 2,
 	D-85748 Garching, 
	Germany    
     }

\maketitle              

\begin{abstract}
How frequent are brown dwarf binaries?
Do brown dwarfs have planets?
Are current theoretical pre-main-sequence evolutionary tracks valid down
to the substellar regime? -- Any detection of a companion to a brown dwarf 
takes us one step forward towards answering these basic questions of star
formation.

We report here on a search for spectroscopic and visual
companions to young brown dwarfs in the Cha\,I star forming cloud.
Based on spectra taken with UVES at the VLT,
we found significant radial velocity (RV) 
variations for five bona-fide and candidate brown dwarfs in Cha\,I.
They can be caused by either a (substellar or planetary)
companion or stellar activity. 
A companion causing the detected RV variations would have 
about a few Jupiter masses. 
We are planning further UVES observations in order to explore
the nature of the detected RV variations. 
We also found that the RV dispersion is only $\sim$2\,km/s
indicating that there is probably no run-away brown dwarf among them.

Additionally a search for companions by direct imaging with the HST and 
SOFI (NTT) has yielded to the detection of 
a few companion candidates in larger orbits.

\end{abstract}

\section{Multiplicity of Brown Dwarfs}
High-precision radial velocity (RV) surveys have brought up more than 60
planetary candidates in orbit around stars (mostly G- and K-type) but only
a few brown dwarf candidates despite the fact
that these surveys are more sensitive to higher masses.
This is referred to as the 'brown dwarf desert'.

The search for fainter companions by direct imaging 
yielded to the discovery of 
seven brown dwarf companions to stars
confirmed by both spectroscopy as well as proper motion. These are
Gl\,229\,B 
(Nakajima et al. 1995, Oppenheimer et al. 1995),
G\,196-3\,B (Rebolo et al. 1998),
Gl\,570\,D (Burgasser et al. 2000),
TWA\,5\,B (Lowrance et al. 1999, Neuh\"auser et al. 2000),
HR\,7329 (Lowrance et al. 2000, Guenther et al. 2001), 
Gl\,417B and Gl\,584C (Kirkpatrick et al. 2000, 2001).
Furthermore GG\,Tau\,Bb (White et al. 1999) and Gl\,86 (Els et al. 2001)
are good candidates of brown dwarf companions to stars.

These detections show that there are at least a few 
brown dwarfs orbiting stars
but what about brown dwarfs having companions themselves?
Up to now there are three brown dwarf binaries known, i.e.
brown dwarf -- brown dwarf pairs: the brown dwarf spectroscopic 
binary PPl\,15 (Basri \& Mart\'{\i}n 1999) and two brown dwarf 
binaries confirmed by both imaging and common proper motion, 
DENIS-P\,J1228.2-1547 (Mart\'{\i}n et al. 1999)
and 2MASSW\,J1146 (Koerner et al. 1999).
Very recently another object, 2MASSs J0850359+105716, 
has been detected as likely brown dwarf binary
(Reid et al. 2001).

There is no planet known orbiting a brown dwarf.
The lowest mass star with a RV planet candidate is the M4-dwarf Gl 876
(Delfosse et al. 1998).
 
Do brown dwarfs have companions at all? 
Planetary or substellar companions around brown dwarfs may form in a 
circumstellar disk but it is also possible that they form by fragmentation 
like low-mass stars. 
Although circumstellar disks around
mid- to late M-dwarfs might be expected to have insufficient mass to
form companions around them, recent observational evidence hints at
significant reservoirs of gas and dust even around these objects
(Persi et al. 2000, Fern\'andez \& Comer\'on, 2001).
There are also indications for the presence of
significant circumstellar material around a few
of the Cha\,I and $\rho$\,Oph 
bona fide and candidate brown dwarfs, which
show IR excess (Comer\'on et al. 2000, Wilking et al. 1999).

\section{Bona fide and candidate Brown Dwarfs in Cha\,I}
The Cha\,I dark cloud is a
site of on-going and/or recent low- and intermediate-mass star formation.
It is one of the most promising grounds for observational projects on
young very low-mass objects, since it is
nearby (160\,pc) and the extinction is low compared to other 
star forming regions.

By means of two H$\alpha$ objective-prism surveys 
12 new low-mass late-type objects (M6--M8) have been detected in the center 
of Cha\,I, named Cha\,H$\alpha$\,1 to 12
(Comer\'on et al. 1999, 2000).
Their masses are below or near the border line separating brown dwarfs 
and very low-mass stars according to comparison with evolutionary tracks by
Baraffe et al. (1998) and Burrows et al. (1997).
Four of them have been confirmed as bona fide
brown dwarfs (Neuh\"auser \& Comer\'on 1998, 1999 and Comer\'on et al. 2000). 
\begin{figure}[t]
\begin{center}
\includegraphics[width=.38\textwidth,angle=270]{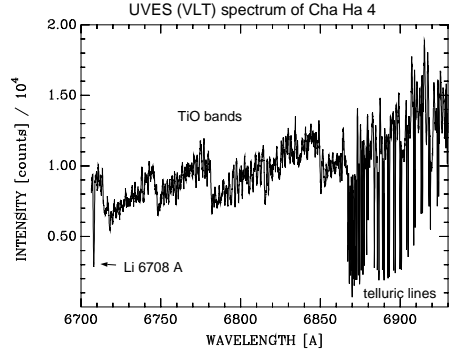}
\end{center}
\caption[]{
UVES Echelle spectrum. 
For clarity only a small part of the total observed spectrum 
(6700\,{\AA} to 10400\,{\AA}) is displayed.
Late M-dwarfs 
exhibit a wealth of spectral features in the red part of the 
wavelength range, which are used for the RV determination by means of
cross correlation. The telluric lines served as
wavelength reference.}
\label{spec}
\end{figure}

\section{The Radial Velocity Survey}

We used UVES, the high-resolution Echelle spectrograph at the VLT
in order to search for companions to the bona fide
and candidate brown dwarfs in Cha\,I.
We took at least two spectra separated by a few weeks of each of 
the nine brightest objects
in the red part of the wavelength range, since the objects are extremely red 
(Fig.\,\ref{spec}).  

The determination of precise RVs requires
the superposition of a wavelength reference  
on the stellar spectrum so that both light beams follow exactly the same path
in the spectrograph.
We used the telluric O$_2$ lines as wavelength reference, 
which are produced by molecular oxygen in the Earth atmosphere 
and show up 
in the red part of the optical spectral 
range (Fig.\,\ref{spec}). It has been shown that they are stable up to
about 20\,m/s (Balthasar et al. 1982, Caccin et al. 1985).
The Iodine cell, often used for extrasolar planet 
searches produces no lines in this wavelength region. 

RVs are determined by cross-correlating plenty of stellar lines of the 
object spectra against a template spectrum and 
locating the correlation maximum.
As a template we used a mean spectrum of a young, cool star also obtained 
with UVES. 
The spectral resolution of the UVES spectra is about 40\,000.
We achieved a velocity resolution of about 200\,m/s for a S/N of 20 
in agreement with the expectations for this S/N (Hatzes \& Cochran 1992).

\subsection{Small radial velocity dispersion}
\begin{figure}[b]
\begin{center}
\includegraphics[width=.3\textwidth,angle=270]{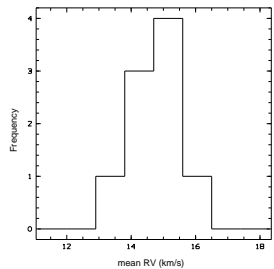}
\end{center}
\caption[]{A histogram of mean RVs of nine bona fide and candidate 
brown dwarfs in Cha\,I clearly depicts
the very small RV dispersion of the studied sample of only $\sim$2\,km/s.
This detection indicates that there is no run-away brown dwarf among 
the sample.}
\label{hist}
\end{figure}
\begin{figure}[b]
\begin{center}
\includegraphics[width=.29\textwidth,angle=90]{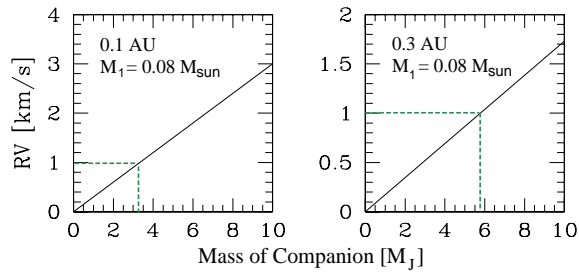}
\end{center}
\caption[]{
The detected RV variations of about 1\,km/s
for five bona fide and candidate brown dwarfs 
in Cha\,I could be caused by planetary companions:
A RV variation of 1\,km/s corresponds to a few Jupiter masses 
depending on the orbital parameters.
}
\label{compmass}
\end{figure}
\begin{figure}[t]
\begin{center}
\includegraphics[width=.35\textwidth,angle=270]{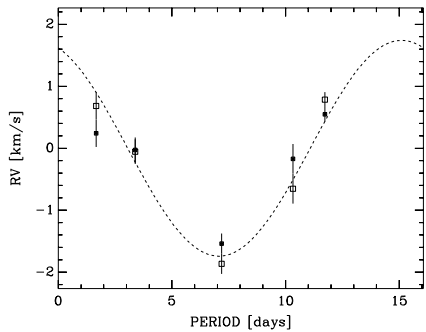}
\end{center}
\caption[]{
Preliminary RV orbit of one of the bona fide and candidate brown
dwarfs.
While five observations are not enough to derive a perfect fit we can 
nevertheless determine an approximate 
M\,sin\,i for the hypothetical companion of 4.8\,M$_{\mbox{\tiny Jup}}$.
}
\label{orbit}
\end{figure}
Neuh\"auser \& Comer\'on (1999) constrained the RV dispersion of the
Cha\,I bona fide and candidate brown dwarfs 
from medium resolution spectra to be 11\,km/s.
Based on the high-resolution UVES spectra 
of nine of the Cha\,I bona fide and candidate brown dwarfs in Cha\,I
we find that the RV dispersion is even smaller, 
namely $\sim$2\,km/s (Fig.\,\ref{hist}).
This finding gives suggestive evidence that there is no run-away brown dwarf 
among them and does not support the formation 
scenario that brown dwarfs are ejected
stellar 'embryos' proposed by Reipurth \& Clarke (2001). 

\subsection{Jupiter mass companions around brown dwarfs?}

The analysis of UVES spectra taken at different times yielded to the 
detection of significant RV variations
for five bona fide and candidate brown dwarfs in Cha\,I. 
They could be caused by reflex motion due to orbiting objects
or by shifting of the spectral line center due to surface 
features (stellar activity).

The detected RV variations are of the order of 1\,km/s. 
If they are caused by companions they
would have masses of a few Jupiter masses depending on the 
orbital parameters as shown in Fig.\,\ref{compmass}.
We found a \emph{preliminary} RV orbit for one of the studied objects	
(Fig.\,\ref{orbit}) with an approximate  
M\,sin\,i for a hypothetical companion of 4.8\,M$_{\mbox{\tiny Jup}}$. 
This shows that the detection of companions with masses of a few
times the mass of Jupiter in orbit around brown dwarfs or very low-mass 
stars is clearly feasible with these data.
An orbiting planet around a brown dwarf has a much larger effect on its
parent object than a planet in orbit around a star and is therefore
easier to detect.
If an absorption cell for the red part of the optical wavelength range
would be available for UVES a RV precision of 3\,m/s might be feasible and
with it the detection of planets with a few Earth masses in orbit around
brown dwarfs.

It would be an interesting finding if the existence of planets around 
the studied bona fide and candidate brown dwarfs in Cha\,I can be confirmed 
since they would be the objects (very low-mass stars or even brown dwarfs)
with the latest spectral types, 
i.e. the lowest masses, harboring planets. 
Furthermore the confirmation of planets around these extremely young objects 
would show that fully formed giant planets can already exist
around low-mass objects that are only a few million years old.
All the up to date known brown dwarf binaries (cp. Sec.\,1) 
are considerably older than the ones in Cha\,I.

\subsection{Stellar spots on brown dwarfs?}

We know from young T~Tauri stars that they exhibit prominent
surface features causing RV variations of the order of 2\,km/s
(Guenther et al. 2000) due to a high level of magnetic activity
of these stars.

It is an outstanding question if brown dwarfs have 
magnetically active photospheres and consequently prominent surface features.
Recent detections of \mbox{X-ray} emission from young brown dwarfs 
may be explained by magnetic activity
(Neuh\"auser \& Comer\'on 1998, Neuh\"auser et al. 1999,
Comer\'on et al. 2000). 
Furthermore a few brown dwarfs have shown indications for periodic
photometric variabilities with periods less than one 
day, which may be caused by spots or weather (Bailer-Jones \& Mundt 2001,
Eisl\"offel \& Scholz, this conference).

We investigated the UVES spectra for hints of magnetic activity.
We detected Ca\,II emission 
for some of the 
bona fide and candidate brown dwarfs in Cha\,I, but
did not find a correlation between the presence of
this emission and 
the RV variations.
Some of the targets have been detected as X-ray emitters 
(Neuh\"auser et al. 1999, Comer\'on et al. 2000) but
there is no correlation between the X-ray luminosity
and the amplitude of the detected RV variations.

Photometric data of the objects will be used to search for rotational
periods. If the RV variations are caused by spots
the brown dwarfs should exhibit photometric variations with the same period.
Furthermore RV monitoring 
may yield useful complementary information on the appearance and 
evolution of cool spots in brown dwarfs and very low-mass stars.

\section{Direct Imaging Campaign}
In a complementary project to the RV survey we are
searching for visual companions to the Cha\,I bona fide and candidate 
brown dwarfs with larger separations
by means of direct imaging.
Images obtained with the HST and SOFI at the NTT yielded to the detection of 
a few companion candidates and are subject of further analysis.

%

\end{document}